# ANALYSIS OF WEB LOGS AND WEB USER IN WEB MINING

L.K. Joshila Grace[1,] V.Maheswari[2], Dhinaharan Nagamalai[3],

[1]Research Scholar, Department of Computer Science and Engineering
joshilagracejebin@gmail.com
[2] Professor and Head,Department of Computer Applications
[1,2]Sathyabama University,Chennai,India
[3]Wireilla Net Solutions PTY Ltd, Australia

## ABSTRACT

*Log files contain information about User Name, IP Address, Time Stamp, Access Request, number of Bytes Transferred, Result Status, URL that Referred and User Agent. The log files are maintained by the web servers. By analysing these log files gives a neat idea about the user. This paper gives a detailed discussion about these log files, their formats, their creation, access procedures, their uses, various algorithms used and the additional parameters that can be used in the log files which in turn gives way to an effective mining. It also provides the idea of creating an extended log file and learning the user behaviour.*

## 1. INTRODUCTION

Log files are files that list the actions that have been occurred. These log files reside in the web server. Computers that deliver the web pages are called as web servers. The Web server stores all of the files necessary to display the Web pages on the users computer. All the individual web pages combines together to form the completeness of a Web site. Images/graphic files and any scripts that make dynamic elements of the site function. , The browser requests the data from the Web server, and using HTTP, the server delivers the data back to the browser that had requested the web page. The browser in turn converts, or formats, the files into a user viewable page. This gets displayed in the browser. In the same way the server can send the files to many client computers at the same time, allowing multiple clients to view the same page simultaneously.

## 2. CONTENTS OF A LOG FILE

The Log files in different web servers maintain different types of information. [6]The basic information present in the log file are

- User name: This identifies who had visited the web site. The identification of the user mostly would be the IP address that is assigned by the Internet Service provider (ISP). This may be a temporary address that has been assigned. There fore here the unique identification of the user is lagging. In some web sites the user identification is made by getting the user profile and allows them to access the web site by using a user name and password. In this kind of access the user is being identified uniquely so that the revisit of the user can also be identified.

- Visiting Path: The path taken by the user while visiting the web site. This may be by using the URL directly or by clicking on a link or trough a search engine.

- Path Traversed: This identifies the path taken by the user with in the web site using the various links.

- Time stamp: The time spent by the user in each web page while surfing through the web site. This is identified as the session.

- Page last visited: The page that was visited by the user before he or she leaves the web site.

- Success rate: The success rate of the web site can be determined by the number of downloads made and the number copying activity under gone by the user. If any purchase of things or software made, this would also add up the success rate.

- User Agent: This is nothing but the browser from where the user sends the request to the web server. It's just a string describing the type and version of browser software being used.

- URL: The resource accessed by the user. It may be an HTML page, a CGI program, or a script.

- Request type: The method used for information transfer is noted. The methods like GET, POST.

These are the contents present in the log file. This log file details are used in case of web usage mining process. According to web usage mining it mines the highly utilized web site. The utilisation would be the frequently visited web site or the web site being utilized for longer time duration. There fore the quantitative usage of the web site can be analysed if the log file is analysed.

## 3. LOCATION OF A LOG FILE

A Web log is a file to which the Web server writes information each time a user requests a web site from that particular server. [7]A log file can be located in three different places:

- Web Servers

- Web proxy Servers

- Client browsers

### 3.1 Web Server Log files

The log file that resides in the web server notes the activity of the client who accesses the web server for a web site through the browser. The contents of the file will be the same as it is discussed in the previous topic. In the server which collects the personal information of the user must have a secured transfer.

## 3.2  Web Proxy Server Log files

A Proxy server is said to be an intermediate server that exist between the client and the Web server. There fore if the Web server gets a request of the client via the proxy server then the entries to the log file will be the information of the proxy server and not of the original user. These web proxy servers maintain a separate log file for gathering the information of the user.

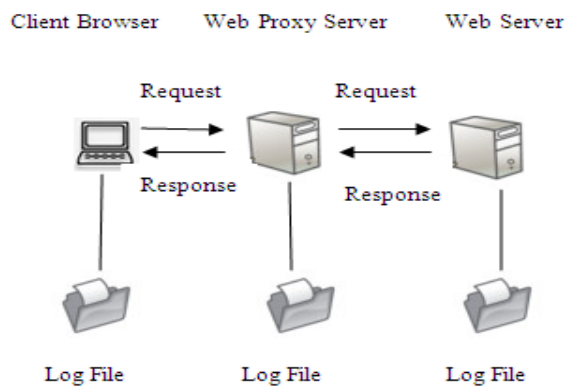

Figure 1.  Web Proxy Server Log files

## 3.3  Client Browsers  Log files

This kind of log files can be made to reside in the client's browser window itself. Special types of software exist which can be downloaded by the user to their browser window. Even though the log file is present in the client's browser window the entries to the log file is done only by the Web server.

# 4  TYPES OF WEB SERVER LOGS

Web Server logs are plain text (ASCII) files and are Independent from the server. [10]There are some Distinctions between server software, but traditionally there are four types of server logs:

- Transfer Log

- Agent Log

- Error Log

- Referrer Log

The first two types of log files are standard. The referrer and agent logs may or may not be "turned on" at the server or may be added to the transfer log file to create an "extended" log file format. [11]The log file entries of Apache HTTP Server Version 1.3 are discussed below:

## 4.1  Error Log

The information that is contained in most error log entries is the message given below

 *[Wed Oct 11 14:32:52 2000] [error] [client 127.0.0.1] client denied by server*

*configuration: /export/home/live/ap/htdocs/test*

When ever an error is occurred while the page is being requested by the client to the web server the entry is made in the error log. The first set of item present in the log entry is the date and time of the message. The second entry lists the severity of the error being reported. The Log Level directive is used to control the types of errors that are sent to the error log by restricting the severity level. The third entry gives the IP address of the client that generated the error. Next is the message itself, which in this case indicates that the server has been configured to deny the client access. The server reports the file-system path of the requested document.

## 4.2 Access Log

The server access log records all requests that are processed by the server. The location and content of the access log are controlled by the Custom Log directive. The Custom Log directive is used to log requests to the server. A log format is specified, and the logging can optionally be made conditional on request characteristics using environment variables.The Log Format directive can be used to simplify the selection of the contents of the logs. This section describes how to configure the server to record information in the access log. here are three log formats considered for access log entries in the case of Apache HTTP Server Version 1.3.They are briefly discussed below:

### 4.2.1 Common Log Format (CLF)

The configuration of the common log format is given below

*LogFormat "%h %l %u %t \"%r\" %>s %b" common CustomLog logs/access_log common*

The log file entries produced in CLF will look something like this:

*127.0.0.1 - frank [10/Oct/2000:13:55:36 -0700] "GET /apache_pb.gif HTTP/1.0" 200 2326*

The entries give details about the client who had requested for the web site to the web server

- 127.0.0.1 (%h) - This is the IP address of the client which made the request to the server.

- - (%l) - The hyphen present in the log file entry next to the IP address indicates that the requested information is not available.

- frank (%u) - The user id of the person requesting the document as determined by HTTP authentication

- [10/Oct/2000:13:55:36 -0700] (%t) -The time format resembles like [day/month/year: hour: minute: second zone]

- "GET /apache_pb.gif HTTP/1.0" (\"%r\") - The request sent from the client is given in double quotes. GET is the method used. apache_pb.gif is the information requested by the client. The protocol used by the client is given as HTTP/1.0

- 200 (%>s) - This is the status code sent by the server. The codes beginning with 2 for successful response, 3 for redirection, 4 for error caused by the client, 5 for error in the server

- 2326 (%b) - The last entry indicates the size of the object returned to the client by the server, not including the response headers. If there is no content returned to the client, this value will be "-"

### 4.2.2 Combined Log Format

The configuration of the combined log format is given below

*LogFormat "%h %l %u %t \"%r\" %>s %b \"%{Referer}i\" \"%{User-agent}i\"" combined*
*CustomLog log/acces_log combined*

The log file entries produced in combined log format will look something like this:

*127.0.0.1 - frank [10/Oct/2000:13:55:36 -0700] "GET /apache_pb.gif HTTP/1.0" 200 2326 "http://www.example.com/start.html" "Mozilla/4.08 [en] (Win98; I ;Nav)"*

The additional parameters that are present in the combined log format are discussed below

- *"http://www.example.com/start.html" (\"%{Referer}i\")* - This gives the site that the client reports having been referred from. (This should be the page that links to or includes /apache_pb.gif).

- *"Mozilla/4.08 [en] (Win98; I ;Nav)" (\"%{User-agent}i\")* - This is the information that the client browser reports about itself to the server.

These entries are made in the log file for a combined log format entry.

### 4.2.3 Multiple Access Logs

Multiple access logs can be created simply by specifying multiple Custom Log directives in the configuration file. In this type of log file there are three files created as access log files containing the details about the client. It is said to be a combination of common log format and combined log format.

The configuration of the multiple access log is given below:

*LogFormat "%h %l %u %t \"%r\" %>s %b" common*
*CustomLog logs/access_log common*
*CustomLog logs/referer_log "%{Referer}i -> %U"*
*CustomLog logs/agent_log "%{User-agent}i"*

The first line contains the basic CLF information, while the second and third line contains referrer and browser information.

Most of the web servers have the same formats being followed for the log file entry.

## 5  STATUS CODES SENT BY THE SERVER

After processing the request of the client in the web server the status code is sent by the web server. There are various status that are exhibited by Apache HTTP Server Version 1.3 is given below:

**1xx Info**

HTTP_INFO – Request received, continuing process

- 100 Continue   – HTTP_CONTINUE
- 101 Switching Protocols -HTTP_SWITCHING_PROTOCOLS
- 102 Processing   – HTTP_PROCESSING

**2xx Success**

HTTP_SUCCESS – action successfully received, understood,  accepted

- 200 OK   – HTTP_OK
- 201 Created   – HTTP_CREATED
- 202 Accepted   – HTTP_ACCEPTED

**3xx Redirect**

HTTP_REDIRECT – The client must take additional action to complete the request xx Info

- 301 Moved Permanently – HTTP_MOVED_PERMANENTLY
- 302 Found   – HTTP_MOVED_TEMPORARILY
- 304 Not Modified  – HTTP_NOT_MODIFIED

**4xx Client Error**

HTTP_CLIENT_ERROR – The request contains bad syntax or cannot be fulfilled

- 400 Bad Request    – HTTP_BAD_REQUEST
- 401 Authorization Required  –HTTP_UNAUTHORIZED
- 402 Payment Required  – HTTP_PAYMENT_REQUIRED
- 404 Not Found   – HTTP_NOT_FOUND
- 405 Method Not Allowed  – HTTP_METHOD_NOT_ALLOWED

**5xx Server Error**

HTTP_SERVER_ERROR – The server failed to fulfill an apparently valid request.

- 500 Internal Server Error – HTTP_INTERNAL_SERVER_ERROR
- 501 Method Not Implemented – HTTP_NOT_IMPLEMENTED
- 503 ServiceTemporarily Unavailable  – HTTP_SERVICE_UNAVAILABLE

- ▪ 505 HTTP Version Not Supported – HTTP_VERSION_NOT_SUPPORTED

Due to space constraints only few status codes are discussed above. These status codes that are sent along with the response data is also entered in the log file.

# 6 OVERVIEW OF WEB MINING

Web mining employs the technique of data mining into the documents on the World Wide Web. The overall process of web mining includes extraction of information from the World Wide Web through the conventional practices of the data mining and putting the same into the website features.

In the web mining process there are three types of mining they are web content mining, Web structure mining, Web usage mining.

## 6.1 Web Structure mining

This involves the usage of graph theory for analyzing the connections and node structure of the website. According to the type and nature of the data of the web structure, it is again divided into two kinds

- Extraction of patterns from the hyperlink on the net: The hyperlink is structural form of web address connecting a web page to some other locations.

- Mining of the structure of the document: The tree like structure gets used for analyzing and describing the XHTML or the HTML tags in the web page.

## 6.2 Web Content mining

In this kind of mining process attempts to discover all links of the hyperlinks in a document so as to generate the structural report on a web page. There are two groups of web content mining strategies. First strategy is to directly mine the content of documents and the second one are those that improve on the content search of other tools like search engines.

## 6.3 Web Usage mining

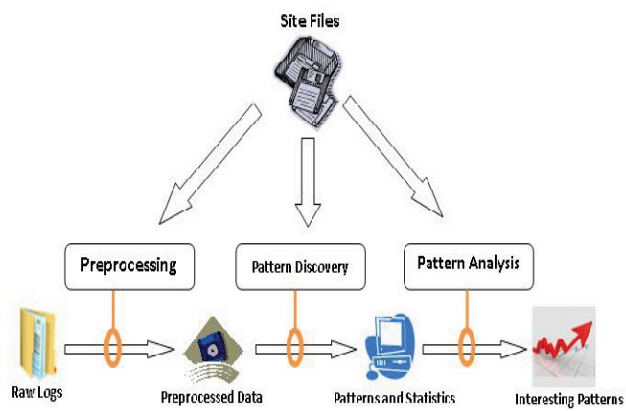

Figure 2. One High level web usage mining process

In the web usage mining process, the techniques of data mining are applied so as to discover the **t**rends and the patterns in the browsing nature of the visitors of the website. There is extraction of the navigation patterns as the browsing patterns could be traced and the structure of the website can be designed accordingly. When it is talked about the browsing nature of the user it deals with frequent access of the web site or the duration of using the web site. This information can be extracted from the log file. Only these log files record the session information about the web pages. [5][8] The fig 2 shows the step wise procedure for web usage mining process

## 7 USING LOG FILE DATA IN WEB USAGE MINING

The contents of the Log files are used in this type of mining. Web usage mining also consists of three main steps:

### 7.1 Preprocessing

The data present in the log file cannot be used as it is for the mining process. [9]There fore the contents of the log file should be cleaned in this preprocessing step. The unwanted data are removed and a minimized lo file is obtained.

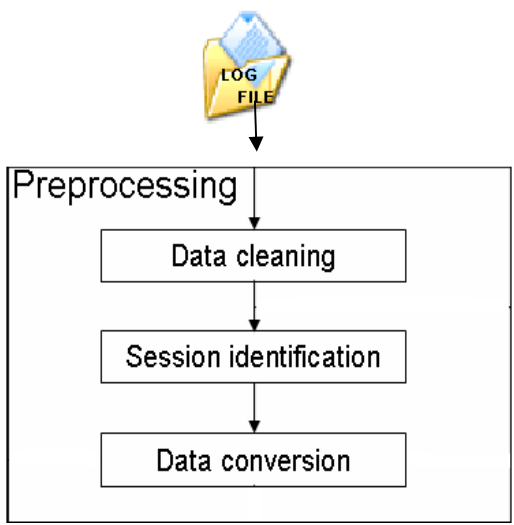

Figure 3.  Pre-processing of Log File

- *Data cleaning:* In this process the entries made in the log file for the unwanted view of images, graphics, Multi media etc., made by the users are removed. Once these data are removed the size of the file is minimized to a greater extent.

- *Session Identification:* This done by using the time stamp details of the web pages. The total time used by each user of each web page. This can also be done by noting down the user id those who have visited the web page and had traversed through the links of the web page. Session is the time duration spent in the web page.

- *Data conversion:* This is conversion of the log file data into the format needed by the mining algorithms.

## 7.2 Pattern discovery

After the conversion of the data in the log file into a formatted data the pattern discovery process is under gone. [8]With the existing data of the log files many useful patterns are discovered either with user id's, session details, time outs etc.,

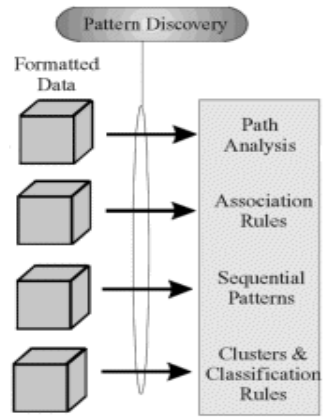

Figure 4.  Pattern Discovery

- *Path analysis:* Graph models are most commonly used for Path Analysis. A graph represents some relation defined on Web pages and each tree of the graph represents a web site. Each node in the tree represents a web page (html document), and edges between trees represent the links between web sites, while the edges between nodes inside a same tree represent links between documents at a web site.

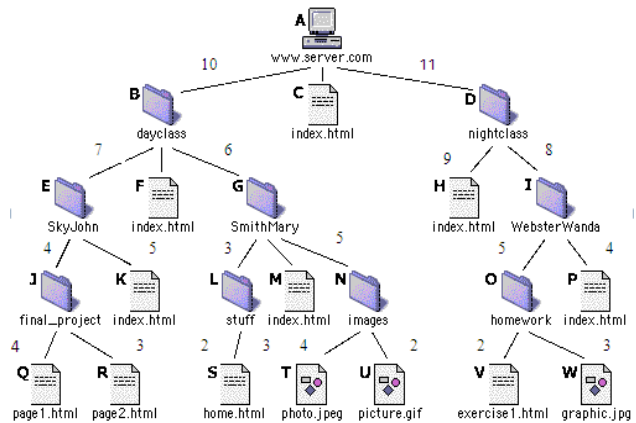

Figure 5.  Path analysis

The Fig 5 shows the access made to a single website. Each node represents the web page within the same web site. Links made is connected to the corresponding nodes and the number of users used each link is also noted with values given between their edges.

- *Association Rules:* This technique is used to predict the correlation of items where the presence of one set of items in a transaction implies the presence of other items. The fig 6 gives the idea of correlation of the same item sets that is with same customer ID or session ID. The example is considered for two different log file details.

Figure 6.  Association rules exhibited by using session ID and Customer ID

- *Sequential Patterns:* sequential patterns discover the user's navigation behaviour.  The sequence of items occurring in one transaction has a particular order between the items or the events. The same sequence of item may re-occur in the same order. For example 30% of the user may under go link in this order "A=>B=>C=>D=>E" where ABCDE corresponds to each web page.

- *Clusters and Classification rule:* This process groups profiles of items with similar characteristics. This ability enhances the discovery of relationships. The classification of Web access logs allows a company to discover the average age of customers who order a certain product. This information can be valuable when developing advertising strategies. But these kind of information involves personal information about the user.

## 7.3 Pattern analysis

This analysis process would eliminate the irrelevant rules or patterns that were generated. They tend to extract the interesting rules or patterns from the output of the pattern discovery process. [4]The most common form of pattern analysis consists of a knowledge query mechanism such as SQL (Structured Query Language) or loads the usage data into a data cube in order to perform OLAP (Online analytical processing) operations. Visualization techniques, such as graphing patterns or assigning colors to different values, can often highlight overall patterns or trends in the data. Various other mechanisms used for mining these patterns are

- *Site Filter:* This technique is used in WEBMINER system. [5]The site filter uses the site topology to filter out rules and patterns that are not interesting. Any rule that confirms direct hypertext links between pages is filtered out.

- *mWAP(Modified Web Access Pattern):* [3]This technique totally eliminates the need to engage the numerous reconstruction of intermediate WAP-trees during mining and considerably reduces execution time.

- *EXT-Prefixspan:* [1]This method mines the complete set of patterns but greatly reduces the efforts of candidate subsequence generation. Prefix –projection process involved in this method substantially reduces the size of projected database.

- *BC-WAPT (Binary coded Web Access pattern Tree):* [2]eliminates recursive reconstruction of intermediate WAP tree during the mining by assigning the binary codes to each node in the WAP Tree

# 8 CREATION OF AN EXTENDED LOG FILE

The log file contents would be even more efficient if it provides the details of the clicks made by the user while visiting the web site. If the user opens a particular web site and does some other work outside the system then it may also be considered as the usage of the web site. The details regarding the clicks made by the user and the time he or she scrolled or did any other operation can also be noted for effective mining of the web usage data.

We shall consider one more situation were the user clicks to open a web site and also works with some other web site in a different browser window. In this situation we can analyze that the user can only read details of only one web site at a time. Then it is understood that the other web site is said to be ideal. But even in this situation the details in the log file would note the input as the web page is being used.

By taking these small differences in the time or the session of the web page being used, still an efficient web mining can be done.

# 9 LEARNING USERS AREA OF INTEREST

Learning the users expectation is a very tedious process. A single word may have different views by different user. If the users area of interest is identified then we can have an efficient mining process. How is this done. If questions are posed to the user it would be a tiring process for a user to answer the question each time he makes a search. Therefore the users interest can be analysed by the first attempt made to open a page. Then the next step done by the miner is to mine the web once again and provide the list of result meant only for the users area of interest. This may inturn minimize the list of options and make the searching process even more effective. This can be done along with analysis of the log files to have utility as one of the factor.

# 10 CONCLUSIONS

The Paper gives a detailed look about the web log file, its contents, its types, its location etc., Added to these information it also gives a detailed description of how the file is being processed in the case of web usage mining process. The various mechanisms that performs each step in mining the log file is being discussed along with their disadvantages. The additional parameters that can be considered for Log file entries and the idea in creating the extended log file is also discussed briefly. The extended work is to combine the concept of learning the users area of interest.

**Authors**

Mrs.L.K. Joshila Grace had her education in Computer Science and Engineering in Jeppiaar Engineering College and Masters degree in Computer Science and Engineering from Sathyabama University Chennai,TN. Presently she is working as a Lecturer in Department of Computer Science and Engineering in Sathyabama University Chennai,TN. Her areas of interest are Data Mining and Warehousing, Artificial Intelligence, Database Management Systems, Java and Web Technologies.

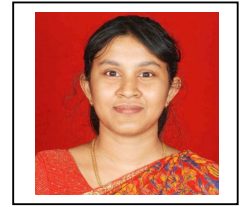